\documentclass{aa}
\usepackage{graphicx}

\newcommand{\be}{\begin{equation}}
\newcommand{\ee}{\end{equation}}

\begin{document}
\thesaurus{12(02.01.2;02.13.2;08.06.2;09.10.1;11.14.1;11.10.1)}

\title{Magnetized accretion-ejection structures}
\subtitle{IV. Magnetically-driven jets from resistive, viscous, Keplerian
  discs}  

\author{Fabien Casse \and Jonathan Ferreira}
\offprints{Fabien.Casse@obs.ujf-grenoble.fr}

\institute{Laboratoire d'Astrophysique de l'Observatoire de Grenoble BP53,
  F-38041 Grenoble cedex 9, France}

\date{Received 1999 July 29/ Accepted 1999 November 24}

\maketitle

\begin{abstract}
We present steady-state calculations of self-similar magnetized accretion
discs driving cold, adiabatic, non-rela\-tivistic jets. For the first time, 
both the magnetic torque due to the jets and a turbulent "viscous"
torque are taken into account. This latter torque allows a dissipation of
the accretion power as radiation at the disc surfaces, while the former
predominantly provides jets with power. 

The parameter space of these structures has been explored. It is
characterized by four free parameters, namely the disc aspect ratio and
three MHD turbulence parameters, related to the anomalous magnetic
diffusivities and viscosity. It turns out that launching cold jets from
thin, dissipative discs implies anisotro\-pic turbulent dissipation. Jets that
asymptotically reach a high Alf\-v\'enic Mach number are only produced by
weakly dissipative discs. 

We obtained general analytical relations between disc and jet quantities
that must be fulfilled by any steady-state model of cold jets, launched from
a large radial extension of thin discs. We also show that such discs cannot
have a dominant viscous torque. This is because of the chosen geometry,
imposing the locus of the Alfv\'en surface.

Some observational consequences of these cold magnetized accretion-ejection
structures are also briefly discussed.

\end{abstract}

\section{Cold jets from Keplerian accretion discs}

Jets of plasma are observed around all young stellar objects of low
mass, some galactic objects  and active galactic nuclei (see the review of
Livio \cite{Livio} and references therein). These jets share common
properties, namely a high  degree of collimation and (at least in some
objects) evidences of interrelations with an underlying accretion disc
(Hartigan et al. \cite{Har}, Serjeant et al. \cite{Sarj}). A category of very
promising models explaining both galactic and extragalactic jets rely on
the interaction between the accretion disc and a large scale magnetic
field (for an alternative view see Lery et al. \cite{LeryII}). The field
extracts angular momentum from the disc, thereby allowing accretion and
drives the jet. This magnetized jet will be naturally self-collima\-ted
provided a large enough current is asymptotically maintained (Chan \&
Henriksen \cite{Chan}, Heyvaerts \& Norman \cite{Hey}).

Following Blandford \& Payne (\cite{BP82}) (hereafter BP82) there have been
numerous studies of magnetized jet dynamics, with prescribed boundary
conditions at the disc surface. As a result, the conviction grew that jets
are indeed magnetized, but the question of their launching remained
open. Indeed, if jets are expected to carry away a substantial fraction of
the disc angular momentum, then a precise treatment of disc-jet
interrelations must be done. In particular, no standard accretion disc
(Shakura \& Sunyaev \cite{Shak}) could be used as a proper boundary condition. 
Hereafter, we call magnetized accretion-ejection structure (MAES), an
object where accretion and ejection are interdependent. 
 
In a series of papers (Ferreira \& Pelletier \cite{FP93}, Ferreira \&
Pelletier \cite{FP95}, Ferreira \cite{F97}, hereafter FP93, FP95 and F97,
respectively), such  structures were investigated using two simplifying
assumptions: (1) jets are cold, i.e. enthalpy plays no role in their
energetics; (2) the magnetic torque due to the jets is dominant. Only
geometrically thin accretion discs (Keplerian) were studied, the turbulent
``viscous'' torque being simply disregarded. In fact, these assumptions are
common to all theoretical works dealing with the connection between
accretion discs and jets (e.g. K\"onigl \cite{Ko},  Wardle \& K\"onigl
\cite{Ward}, Li \cite{LiI}, Li \cite{LiII}). 

Thanks to a self-similar formulation allowing to take into account all
dynamical terms, a smooth transition between the resistive accretion disc
and the ideal MHD jet was achieved. The necessary conditions for steadily
launching cold jets were thereby obtained. It was shown that an
equipartition field is required, i.e. a magnetic energy density close to
the thermal energy density (on the disc midplane). Moreover, the vertical
gradient of plasma pressure is the only force that can gently expells off
matter in the resistive region, against both magnetic and gravitational
compression. Above the disc (at typically two scale-heights), matter flows
along magnetic surfaces and enthalpy is indeed negligible with respect to
other energies. By this process, from 1\% to 10\% only of accreted
mass can be ejected. Since the transition from accretion to ejection occurs
at the disc surface, the parameter space is highly sensitive to any
approximation made on the disc vertical equilibrium (see discussion in
F97).

The energetic budget of a cold MAES is quite significantly different from
that of a standard disc. Indeed, the available mechanical power is shared
by radiation released at the disc surfaces and an outward MHD Poynting flux
that powers the jets. But in a thin disc, all the available energy is
stored as rotation. Therefore, the dominant torque dictates the form and
amount of dissipation that occurs to that energy. While the viscous torque
produces only dissipation, hence a local heating source, the magnetic torque
produces both another reservoir of energy (outward MHD Poynting flux) and
a local dissipation (Joule heating). As a consequence, a stationary MAES
with a dominant magnetic torque is such that almost all the liberated
power feeds the jets (FP95).   

This quite obvious result has strong observational consequences. Whenever
jets are observed and believed to be disc-driven, then one should observe
as well a lack of emission from the inner parts of the disc. This lack
could be interpretated as a ``hole'' (Rutten et al. \cite{rutten}),
corresponding to the radial extension of the magnetized disc (cold
MAES). However, remember that this is a direct consequence of our
assumption (2). Would it then be possible to get both a significant
emission from the disc and jets? 

The aim of this paper is to relax the assumption of a dominant magnetic
torque, keeping however the approximation of cold jets (thin or even slim
discs, but without a hot corona). In Section 2, we write down the set of
local MHD equations governing the whole structure, present all relevant
parameters and provide general analytical results on disc-jet
interrelations. In the following Section, we describe our numerical method to 
solve the problem and display global solutions, from the disc midplane to
the super-Alfv\'enic jet regime. We explore the parameter space for
magnetically-driven jets from resistive and viscous discs and compare our
results to other models. We then conclude by summarizing our findings in
Section 4.

\section{Cold, non-relativistic MAES}

\subsection{General MHD equations}

Our MAES is composed of an accretion disc settled around a central mass
$M_*$ (compact object or young star) and threaded by a large-scale magnetic
field. The presence of such a field could be explained by two different
phenomena: advection of interstellar magnetic field and/or local magnetic
field generation by a disc dynamo (Pudritz \cite{Pudr}, Khanna \& Camenzind
\cite{Khan}, Torkelsson \& Brandenburg \cite{Brand}, R\"udiger et
al. \cite{Rud}). Although a quadrupolar topology often arises from these
studies, it must be kept in mind that (i) they are kinematic, hence not
taking into account the magnetic feedback on the plasma motion (see
Yoshizawa \& Yokoi \cite{Yosh}) and (ii) neglect a possible primordial field
that could have been advected along with the flow. Thus, it is quite
difficult to infer from these works what would be the final magnetic
topology in a realistic accretion disc. Moreover, it has been shown that
under restricted conditions, a quadrupolar topology could produce jets from
a Keplerian accretion disc, although much less powerful than those from a
bipolar topology (see F97, Appendix A). We will thus assume here a bipolar
topology for the disc magnetic field.

The whole MAES is described in a non-relativistic framework, so there is a
limitation of our calculations when applied to the central parts of an
accretion disc around compact objets. However, we are primarily interested
in the interrelations between the disc and its jets. Thus, we only need to
verify that matter remains sub (or even midly) relativistic close to the
disc (i.e., until the Alfv\'en point). We neglect the disc self-gravity
with respect to the gravitational field produced by the central mass and
look for axisymmetric, stationnary solutions.

Thanks to axisymmetry, the vectorial quantities expressed in cylindrical
coordinates ($r$, $\phi$, $z$), can be decomposed into poloidal and
toroidal components, e.g. ${\bf u}={\bf u}_p + \Omega r {\bf e}_{\phi}$ and
${\bf B}={\bf B}_p + B_{\phi} {\bf e}_{\phi}$. A bipolar configuration
allows us to describe the poloidal magnetic field as
\be 
{\bf B}_p = \frac{1}{r}\nabla a \times {\bf e}_{\phi}  \ ,
\ee 
\noindent where $a (r,z)$ is an even function of $z$ and $a=constant$
describes a surface of constant magnetic flux ($a=rA_{\phi}$, $A_{\phi}$
being the toroidal component of the potential vector). The following set of
equations describes a non-relativistic MAES (FP93): 
\begin{itemize}
\item{Mass conservation 
    \be 
    \nabla.\rho{\bf u}= 0
    \label{1}
    \ee}
\item{Momentum conservation 
    \be 
    \rho{\bf u}.\nabla{\bf u}=-\nabla P-\rho\nabla\Phi_G + {\bf J\times
      B} + \nabla.{\bf T} 
    \label{2}
    \ee}
\item{Ohm's law and toroidal field induction
    \begin{eqnarray}
      \eta_mJ_{\phi} {\bf e}_{\phi} &=&{\bf u_p\times B_p}\label{3}\\
      \nabla.(\frac{\nu'_m}{r^2}\nabla rB_{\phi}) & = &
      \nabla.\frac{1}{r}(B_{\phi}{\bf u}_p-{\bf B}_p\Omega r) \label{4}
    \end{eqnarray}}
\end{itemize}
\noindent where $\rho$ is the density of matter, $P$ the thermal pressure,
$\Phi_G=-GM_*/(r^2+z^2)^{1/2}$ the gravitational potential, 
${\bf J}=\nabla\times{\bf  B}/\mu_o$ the current, ${\bf T}$ the ``viscous''
stress tensor (Shakura \& Sunyaev \cite{Shak}), $\nu_m= \eta_m/\mu_o$ and
$\nu'_m$ the (anomalous) poloidal and toroidal magnetic diffusivities. All
transport coefficients appearing in the above set of equations, namely $\nu_m$,
$\nu'_m$ and $\nu_v$ (``viscosity'', contained in ${\bf T}$), are assumed
to be of turbulent origin. 

Energy conservation is not self-consistently solved here (see the
discussion in FP95). Since jets are cold, the precise way energy is
transferred from the disc midplane to the jet plays no dynamical role. We can
therefore freely specify the temperature profile. While previous works
assumed isothermal magnetic surfaces, we will here use an adiabatic equation
\be 
P= K(a)\rho^{\gamma} \ ,
\label{polyt}
\ee 
\noindent where $\gamma=\frac{5}{3}$ for a monoatomic gas and $K(a)$
(related to the specific entropy) is conserved along each field
line. Finally, in order to close the system, we use the perfect gas law 
\be 
P = nk_B T
\label{perfgas}
\ee 
\noindent where $T$ is the temperature of the plasma, $k_B$ the Boltzmann
constant and $n=\rho/m_p$ ($m_p$ being the proton mass).

\subsection{Relevant disc parameters}

Taking into account all dynamical effects appearing in Eq.~(\ref{1}) to
(\ref{perfgas}) is costly, for it results in quite a large number of
dimensionless parameters. However, thanks to various constraints, only four
of them are really free (but the parameter space of a cold MAES
remaining to be spanned is large, see Sect.~3.3). Below, we list all the
relevant parameters of a cold MAES.

The strength of the magnetic field in the disc is measured by 
\be 
\mu = \left.\frac{B^2}{\mu_oP}\right|_{z=0} \ .
\ee 
\noindent This parameter cannot be much smaller than unity for the disc
would become prone to shearing instabilities (Balbus \& Hawley \cite{Balb}). In
the present work however, we always find $\mu$ of order unity (rough
equipartition). 

The efficency of ejection, defined as 
\be 
\xi = \frac{d \ln \dot{M}_a}{d \ln r} 
\ee 
\noindent is the key parameter linking accretion to ejection (mass
conservation).  For $\xi=0$ we obviously have a standard disc, whereas a
typical value for cold MAES is $10^{-2}$. This parameter is strongly
constrained by both the disc vertical equilibrium (minimum value of $\xi$)
and the steady production of super-Alfv\'enic jets (maximum value, F97).
This local ejection efficency is actually related to the magnetic flux
distribution. When the magnetic flux scales as a power law of index $\beta$
(i.e., $\beta = d\ln a/d \ln$ $r_o$ is a constant), the following scaling
must be fulfilled in a Keplerian disc: $\beta= 3/4 + \xi/2$ (note that
BP82 used $\beta=3/4$).

The (effective) magnetic Reynolds number, defined as 
\be 
{\cal R}_m=\left.\frac{ru_r}{\nu_m}\right|_{z=0}=2-\beta +
\frac{r^2}{\beta l^2} 
\ee 
\noindent where 
\[ \frac{a(r,z=0)}{l^2} = -\left . \frac{\partial^2 a}{\partial
    z^2}\right|_{z=0} \ , \]
\noindent is a direct measure of the bending of the magnetic field lines at
the disc midplane (where, from Eq.~(\ref{3}), one gets the right handside
of the above equation). In previous studies, ${\cal R}_m$ was always
of order $\varepsilon^{-1}$, hence providing a large amount of toroidal
current at the disc midplane (FP95, F97). This current induces a large
radial component of the magnetic field, such that $B_r \ga B_z$ at the disc
surface. This is the required bending for magnetically launching cold
material (BP82). However, another situation with a much smaller ${\cal
R}_m$ ($> 2 - \beta$) could still lead to cold ejection, provided there is
an extra source of toroidal current at the disc surface.

The ratio of the magnetic torque due to the jet to the turbulent
``viscous'' torque at the disc midplane, namely
\be 
\Lambda= \left . \frac{({\bf J\times B})_{\phi}}{(\nabla.{\bf T})_{\phi}}
\right|_{z=0} \ ,
\ee 
\noindent could, in principle, range from zero (standard disc) to
infinity (all previous studies on magnetized accretion disc have neglected
the viscous torque). Note that by ``viscous'' torque, we mean a radial
transport of angular momentum. This ``viscous'' transport could originate
from MHD instabilities, i.e. fluctuations of small scale magnetic fields
(Stone et al. \cite{stone}).  

The toroidal magnetic field at the disc surface is measured by 
\be 
q = -\frac{h}{B_o}\left.\frac{\partial B_{\phi}}{\partial z}\right|_{z=0} 
\ee
\noindent The shearing of the magnetic configuration is related to the current
density flowing at the disc midplane and thus depends on the global
electric circuit. This parameter was found to be of order unity for
strongly bent poloidal field (i.e., $B_{\phi} \sim B_r \sim B_z$ at the disc
surface, F97). 

All above parameters are determined by physical constraints. Below, we
present the parameters that remain free, one geometrical and three related
to the MHD turbulence ($\nu_m, \nu'_m, \nu_v$).

{\it (i)} The first parameter is the disc aspect ratio
\be 
\varepsilon = \frac{h(r)}{r} 
\ee 
\noindent where $h(r)$ is the local disc vertical (pressure) scale height. 
Having no energy equation, this parameter remains free and $\varepsilon$ 
is a constant all over the magnetized disc (FP93). With our treatment, we
are able to keep all terms in the equations. Therefore, we do not use the
usual approximation $\varepsilon \ll 1$. In fact, solutions will be
obtained for thin ($\varepsilon =0.01$) to slim ($\varepsilon =0.1$) discs.

{\it (ii)} The second parameter measures the strength of the MHD turbulence 
\be 
\alpha_m = \left.\frac{\nu_m}{V_A h}\right|_{z=0} 
\ee 
\noindent where $V_A$ is the Alfv\'en speed. If one requires stability
against resistive instabilities (naively, $\tau_{\nu}= h^2/\nu_m \la \tau_A =
h/V_A$), one gets $\alpha_m$ of order unity.

{\it (iii)} The third parameter is the magnetic Prandtl number 
\be 
{\cal P}_m = \left . \frac{\nu_v}{\nu_m}\right|_{z=0} 
\ee 
\noindent which measures the ratio of the ``viscosity'' to the poloidal
magnetic diffusivity at the disc midplane. This parameter plays an
important role in the angular momentum equation, which writes at the disc
midplane 
\be 
{\cal P}_m (1 + \Lambda) = {\cal R}_m  \ .
\label{eq:lien}
\ee 
\noindent Our usual understanding about turbulence would impose 
${\cal P}_m$ of order unity (Pouquet et al. \cite{pouq76}). If this remains
true in a MAES, there is a strong link between the dominant torque
($\Lambda$) and the magnetic field curvature (${\cal R}_m$): strongly bent
fields  (${\cal R}_m \sim \varepsilon^{-1}$) extract all disc angular
momentum. The global energy budget of a cold MAES (with $\varepsilon \ll 1$)
reads  
\be
P_{lib} \simeq 2 P_{rad} \; + \; 2 P_{MHD} \ ,
\ee
\noindent where $P_{lib}$ is the liberated mechanical power, $P_{rad}$ the
disc luminosity and $P_{MHD}$ the outward MHD Poynting flux. The liberated
power is $P_{lib} \la P_{acc} \equiv GM_*\dot M_{ae}/2 r_i$, where $r_i$ is
the disc inner radius and $\dot M_{ae}$ is the accretion rate at the disc 
outer edge $r_e$ (FP95). Using the disc angular momentum conservation gives
\begin{eqnarray}
\frac{2P_{rad}}{P_{lib}} & \simeq & \frac{1}{1 + \Lambda} \nonumber \\
\frac{2P_{jet}}{P_{lib}} & = & \frac{\Lambda}{1 + \Lambda} \ .
\end{eqnarray}
\noindent Thus, a stationary MAES with ${\cal P}_m \sim 1$ and a dominant
magnetic torque is such that only a fraction $h/r$ of the total power is
released as radiation. However, for the purpose of a general investigation,
we will keep ${\cal P}_m$ free in this paper.  

{\it (iv)} The last parameter measures the anisotropy of the magnetic
diffusivities 
\be 
\chi_m = \left .\frac{\nu_m}{\nu'_m}\right|_{z=0} 
\ee 
\noindent at the disc midplane. An isotropic turbulent dissipation
would be described with $\chi_m$ of order unity. However, we expect the
leading instabilities triggered inside the MAES to produce enhanced
dissipation of the toroidal field, thus providing $\chi_m<1$.

\subsection{Launching jets from Keplerian discs}
 
In order to eject matter from the disc, a necessary condition is that the
magnetic torque changes its sign at the disc surface. Indeed, it brakes the
matter in the disc midplane and both angular momentum and energy are stored
in the magnetic field. Above the disc, angular momentum and energy must be
given back to matter for launching a jet, namely $({\bf J\times B}).{\bf
  e}_{\phi} > 0$ at $z=h$. The toroidal component of the Lorentz force is
mainly governed by the behaviour of the radial current density, which is in
turn controled by the induction equation (5). Integrated vertically, this
equation gives
\be 
\eta'_m J_r = \eta'_o J_o + r\int_{0}^{z}dz{\bf B}_p.\nabla\Omega -
B_{\phi}u_z \ .
\label{induc} 
\ee 
\noindent The last term (field advection) is negligible in the resistive
disc but will exactly balance the differential rotation effect in the ideal
MHD region. The change of sign of the magnetic torque implies that 
the radial current density {\it must} decrease on a disc scale height. After
some approximations\footnote{see Appendix B of F97. In order to obtain a
  generalization of his expressions, replace his condition $\Gamma\simeq
  1$ by $\Gamma \simeq {\cal R}_m\varepsilon$.}, the necessary condition
for launching a jet is found to be
\be 
\Lambda = \frac{3\chi_m}{\alpha_m^2\varepsilon {\cal P}_m} \ .
\label{Lambda}
\ee 
\noindent This very important relation does not allow much
freedom in thin discs. Indeed, for conventional values of the turbulence
parameters ($\alpha_m \sim {\cal P}_m \sim \chi_m \sim 1$), one gets
$\Lambda \sim \varepsilon^{-1}$, namely a dominant magnetic torque. To
enforce comparable torques ($\Lambda \sim 1$), one then must ask for either
an anisotropic turbulence ($\chi_m <1$) or a large ${\cal P}_m$ (or
both). A fortiori, launching cold jets with $\Lambda\ll 1$ seems almost
impossible. Cold jets have a tremendous influence on the disc
structure. One useful quantity to evaluate is the ratio $\sigma$ 
of the MHD Poynting flux to the kinetic energy flux,
\be
\sigma_{SM} \simeq \frac{\Lambda}{\xi (1+\Lambda)} \ ,
\ee
\noindent measured here right above the disc (at the slow magnetosonic
point). It shows that, unless $\xi$ is of order unity or $\Lambda \sim
\xi$, the magnetic field completely dominates matter at the disc surface
(i.e., $\sigma_{SM} \gg 1$). 

To summarize, $\Lambda$ is constrained by the very existence of a jet,
${\cal R}_m$ by the disc angular momentum conservation and $q$ is an
explicit functions of the other parameters:
\be
q=\frac{\alpha_m}{2}{\cal
  R}_m\varepsilon\delta\mu^{-1/2}\frac{\Lambda}{1+\Lambda} \ ,
\ee
\noindent with $\delta = \Omega_o/\Omega_K$, ratio of the angular velocity
at the disc midplane to the Keplerian rate $\Omega_K$ (see the exact
expression of $\delta$ in Eq.~(\ref{delta})). The two
remaining disc parameters, $\xi$ and $\mu$, are constrained by the smooth
crossing of two critical points (slow magnetosonic and Alfv\'en, see
FP95). Thus, the present model has only four free parameters. In the
next section, we show the strong links between cold jet parameters and those
describing the disc. In practice however, we will impose a set of disc
parameters ($\varepsilon, \alpha_m, {\cal P}_m, \chi_m$) and obtain a
posteriori the jet parameters.

\subsection{Relevant jet parameters}

The jet is the region where the transport coefficients are equal to
zero, namely the ideal MHD medium above the disc. In this regime, mass and
flux conservations combined with Ohm's law (\ref{3}) provide
\be  
{\bf u}_p = \frac{\eta (a)}{\mu_o\rho}{\bf B}_p
\label{MHDI1}
\ee 
\noindent where $\eta(a)=\sqrt{\mu_o\rho_A}$ is a constant along a magnetic
surface and $\rho_A$ is the density at the Alfv\'en point, where the
poloidal velocity reaches the poloidal Alfv\'en velocity. The induction
equation (\ref{4}) becomes 
\be 
\Omega_*(a) = \Omega -\eta\frac{B_{\phi}}{\mu_o\rho r} \ ,
\label{MHDI2}
\ee 
\noindent where $\Omega_*(a)$ is the rotation rate of a magnetic
surface (very close to the Keplerian value). In the jet, plasma flows along
a magnetic surface with a total velocity ${\bf u} = (\eta/\mu_o\rho) 
{\bf B} + \Omega_*r{\bf e}_{\phi}$ not parallel to the total field. 

The angular momentum conservation in the jet simply writes
\be 
\Omega_*r^2_A = \Omega r^2 - \frac{rB_{\phi}}{\eta}
\label{MHDI3}
\ee 
\noindent where $r_A$ is the Alfv\'en radius. Above the disc, the turbulent
torque vanishes and only remains a magnetic accelerating torque.

The projection of the momentum conservation equation a\-long a magnetic
surface provides the Bernoulli equation 
\be
\frac{u^2}{2} + H + \Phi_G -\Omega_*\frac{rB_{\phi}}{\eta} = {\cal E}(a) +
\Omega_*^2r^2_A = E(a) \ ,
\label{Bern}
\ee 
\noindent where $E(a)$ is the constant specific energy carried by the
jet, $H$ is the specific enthalpy defined as $H= \gamma P/(\gamma -1)\rho$
in the adiabatic case. Enthalpy is negligible for ``cold'' plasma 
ejection at the disc surface.   

The shape of the magnetic surface is given by the jet transverse
equilibrium, namely the Grad-Shafranov equation that can be written in the
following form 
\be (1-m^2)J_{\phi} = J_{\lambda} + J_{\kappa} \ ,
\label{Grad}
\ee 
where the sources of current are 
\begin{eqnarray}
J_{\lambda} & = &\rho r\left(\frac{d{\cal
E}}{da}+(1-g)\Omega_*r^2\frac{d\Omega_*}{da} +
g\Omega_*\frac{d\Omega_*r^2_A}{da}\right)\nonumber\\ 
J_{\kappa} & = & r\frac{B^2_{\phi}-m^2B^2_p}{2\mu_o}\frac{d\ln\rho_A}{da} +
m^2\frac{\nabla a}{\mu_or}.\nabla\ln\rho\nonumber
\end{eqnarray}
\noindent Here, $m^2=u^2_p/V^2_{Ap}$ is the Alfv\'en Mach number and
$g=1- \Omega/\Omega_*$ (Pelletier \& Pudritz \cite{PP92}, hereafter PP92). 

We choose to use the usual jet parameters as defined by BP82. Equation
(\ref{MHDI1}) allows to define a mass load parameter
\be 
\kappa = \eta \frac{\Omega_o r_o}{B_o} \ , 
\ee 
\noindent where the subscript $o$ defines quantities at the disc midplane.
This dimensionless parameter describes the mass flux per magnetic flux unit
($d \dot M_j/da= 2 \pi \eta/\mu_o$), thus it is constrained by the disc
vertical equilibrium.

The total specific angular momentum, defined by 
\be 
\lambda = \frac{\Omega_* r^2_A}{\Omega_o r^2_o}\simeq \frac{r^2_A}{r^2_o}  
\ee 
\noindent provides a measure of the magnetic lever arm acting on the disc,
i.e. a measure of the Alfv\'en radius $r_A$. In the case where all the MHD
Poynting flux is converted into jet kinetic power, the magnetic lever arm
parameter uniquely determines the jet terminal velocity
\be
u_{p,max}  =  \Omega_or_o(2\lambda - 3)^{1/2} \ .
\label{upm}
\ee

The asymptotic jet behaviour is strongly influenced by another normalized
quantity. This last jet parameter (``fastness'' parameter) is a direct 
measure of how fast the magnetic rotator is (Michel \cite{Michel}, PP92,
F97, Lery et al. \cite{lery}) 
\be
\label{defome} 
\omega_A = \frac{\Omega_* r_A}{V_{Ap,A}} \ .
\ee 
\noindent PP92 showed that super-Alfv\'enic jets require $\omega_A$
bigger than unity. This subtle parameter links rotation of the magnetic surface
to poloidal motion. In our calculations, it is intimately related to the 
MHD power still available at the Alfv\'en surface. For $\omega_A$ bigger
than but close to unity (or $g_A < 1/2$), matter reaches the Alfv\'en
surface at the expense of all the MHD Poynting flux (no more current
available) but recollimation ta\-kes place just after it. For larger
$\omega_A$ (typically bigger than 1.4, or $g_A > 1/2$), a large amount of  
current is still available and the jet propagates much farther away before
recollimation occurs (F97). The ``fastness'' parameter $\omega_A$
determines also the magnetic geometry at the Alfv\'en surface, both pitch 
($\arcsin$ $(-B_{\phi}/B_p)$) and opening ($\theta= \arccos (B_z/B_p)$)
angles. The larg\-er $\omega_A$, the larger these angles (centrifugal effect
stronger than the hoop stress), providing a subsequent large widening of
the jet. The fastness parameter $\omega_A$ is directly proportional to the
toroidal field at the disc surface $q$ (see below). This parameter did not
explicitely appear in BP82, but it was hidden in the initial angle
of field lines at the disc surface (the larger angle, the larger
$B_{\phi}$).  

In usual treatments of MHD jets, where the disc is a mere boundary
condition, $\kappa$ is fixed by the regularity condition at the slow point
whereas $\lambda$ is fixed by the Alfv\'enic condition for a given
$\omega_A$. In our treatment, all these cold jet parameters are completely
fixed by the set of disc parameters. Using ideal MHD equations, mass
conservation and angular momentum conservation, we get 
\begin{eqnarray}
  \lambda &=& 1+\frac{\sigma_{SM}}{2} \simeq  1 + \frac{\Lambda}{2\xi(1 +
    \Lambda)} \nonumber \\ 
  \kappa &=&  \frac{q}{\lambda - 1} \simeq  \alpha_m({\cal P}_m\varepsilon +
  \frac{3\chi_m}{\alpha_m^2})\delta\xi\mu^{-1/2}\label{qlamb}\\
  \omega_A & \simeq & q\frac{\lambda^{3/2}}{\lambda-1}\frac{\sin
    (\phi_A-\theta_A)}{\sin\phi_A} \nonumber 
\end{eqnarray}
\noindent where $\phi_A$ is the angle between the Alfv\'{e}n surface and
the vertical axis, and $\theta_A$ is the opening angle estimated at
the Alv\'{e}nic transition. This angle $\theta_A$ is implicitly determined
by the Grad-Shafranov equation (see Appendix A). The angle $\phi_A$ is imposed
by the chosen geometry of the Alfv\'{e}n surface. Note that
this last expression is valid for conical Alfv\'{e}n surfaces (see Appendix
B), a geometry which arises naturally when the above parameters vary slowly
across the jet. Rewriting the last expression as 
\begin{equation}
  \label{omegaa}
  \omega_A \sim
  \alpha_v\varepsilon\Lambda \frac{\lambda^{3/2}}{\lambda-1}\frac{\sin
    (\phi_A-\theta_A)}{\sin\phi_A} \ ,
\end{equation}
\noindent where $\alpha_v$ is the usual alpha parameter for turbulent
viscosity (Shakura \& Sunyaev \cite{Shak}), and using the necessary
condition for super-Alfv\'{e}nic jets ($\omega_A > 1$) provides 
\begin{equation}
  \Lambda > \frac{\lambda-1}{\lambda^{3/2}} \frac{1}{\alpha_v \varepsilon}
  \ . 
\end{equation}
This is a strong constraint on the underlying accretion process. A dominant
viscous torque ($\Lambda < 1$) would then require either (1) $\alpha_v$
smaller than unity but huge magnetic lever arms, namely $\lambda >
\varepsilon^{-2}$, or (2) $\alpha_v > \varepsilon^{-1}$. Whether the last
condition is clearly unphysical, the former remains open at this stage.

\begin{figure}
   \includegraphics[angle=-90,width=\columnwidth]{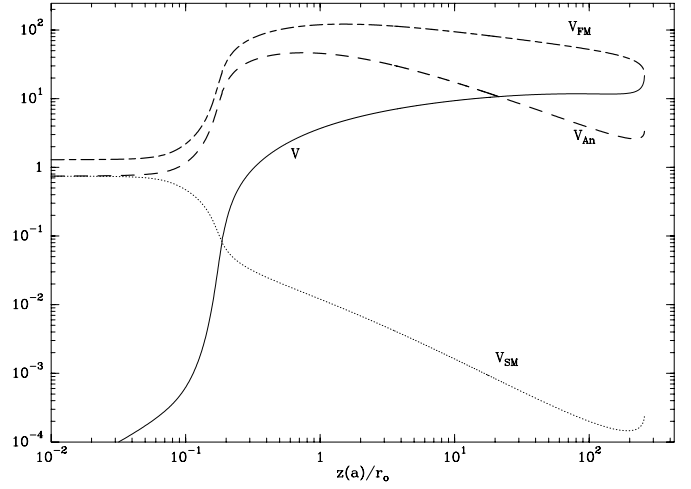}
     \caption{Characteristic velocities along a magnetic surface,
       normalized to the local disc sound speed $\Omega_K h_o$: slow and
       fast magnetosonic velocities $V_{SM}$ and $V_{FM}$, critical plasma
       velocity $V = {\bf u}.{\bf n}$ (${\bf n}$ is the relevant direction
       of wave propagation, FP95) and the Alfv\'en velocity $V_{An}$
       (note that $V= V_{An}$ is equivalent to $u_p = V_{Ap}$). These curves
       describe the MAES presented in Fig.~\ref{jetcar}. However, the
       behaviour is common to all self-similar, cold jets from Keplerian 
       discs obtained so far. Note that the jet is super-fast in the usual
       sense ($u_p > V_{FM}$, F97).}
     \label{jetvel}
\end{figure}

\section{Global solutions from disc to jets}

\subsection{Numerical approach}

\subsubsection{The self-similar ansatz}

In order to have a global solution for a MAES, we need to solve the whole
set of MHD equations (\ref{1}) to (\ref{perfgas}). Because these equations
are highly non-linear partial differential equations, we use a variable
separation method that greatly simplifies the resolution. Self-similarity
allows us to write all the quantities like 
\be 
Q(r,z) = Q_e\left(\frac{r}{r_e}\right)^{\alpha_Q}f_Q(x) 
\ee 
\noindent where $x=z/h(r)$ and $r_e$ is the outer radius of the magnetized
disc. By doing this, we obtain a set of ordinary differential equations
acting on $f_Q$ where we can consider all terms of each equation. The
values of the coefficents $\alpha_Q$ are given by FP93. The magnetic flux
is written  
\be 
a(r,z) = a_e\left(\frac{r}{r_e}\right)^{\beta}\psi(x) 
\ee 
\noindent To obtain the variation of any quantity $Q$ along a magnetic surface,
we just have to write 
\be 
Q(x) = Q_o\psi^{-\alpha_Q/\beta}(x)f_Q(x) 
\ee
\noindent where $x$ must be understood here as a curvilinear abscissa along
the magnetic surface. In fact, we find more convenient to choose as a
curvilinear variable $z(a)/r_o = \varepsilon x \psi^{-1/\beta}(x)$, since it
directly provides a physical scale once $r_o$ is chosen.

\subsubsection{Turbulent transport coefficients}

In the self-similar framework, only the vertical variation of the turbulent
transport coefficents must be prescribed. Since these anomalous coefficents
are expected to arise from an MHD turbulence triggered inside the disc, we
assume that the magnetic diffusivities $\nu_m$ and $\nu_m'$ decrease on a
disc scale-height. We therefore used a Gaussian law for their vertical
profile. 

The same issue appears for the turbulent ``viscous'' torque, which is
responsible for the radial transport of angular momentum. Following the
above arguments, the corresponding turbulent viscosity $\nu_v$ would
also decrease on a disc scale-height. However, we must take care to
conserve the total disc angular momentum. For this purpose, we write the
angular momentum conservation equation as 
\be 
\nabla.( \rho\Omega r^2 {\bf u}_p - \frac{rB_{\phi}}{\mu_o}{\bf B}_p
-r\vec{\tau}) = 0 
\ee 
\noindent where we used
\[\frac{1}{r}\nabla.(r\vec{\tau}) = (\nabla.{\bf T}).{\bf  e}_{\phi} 
\ \ \ \mbox{ and }\ \ \  \vec{\tau}=\left( \begin{array}{c} 
    \tau_{r}\\ 0 \\ \tau_{z}
  \end{array}  \right) \]
\noindent MHD discs launching cold jets have comparable radial and
vertical gradients of the angular velocity $\Omega$, hence providing
comparable components $T_{r\phi}$ and $T_{z\phi}$ of the turbulent stress
tensor. Thus, prescribing the vertical profile of the poloidal
components of $\vec{\tau}$ (instead of the torque itself or the viscosity),
insures the conservation of the total angular momentum. Because the
turbulent {\it torque} must be an even function of $x$, we use a $\exp (-x^4)$
law intead of a simple Gaussian law. The resulting turbulent torque 
remains always negative in the disc and decreases on a disc scale height.

\subsection{Numerical results}

With a given set of disc parameters ($\varepsilon, \alpha_m, {\cal P}_m,
\chi_m$), we integrate the set of ordinary differential equations step by
step with respect to the self-similar variable $x= z/h(r)$. The transition
between the resistive, viscous disc and the ideal MHD medium occurs
smoothly above the disc surface.

The determinant appearing in the set of ideal MHD equations vanishes when
the critical plasma velocity is equal to the phase velocity of a typical
wave of the medium: slow magnetosonic (SM), Alfv\'en (A) and fast
magnetosonic (FM) waves. As already shown by many authors (e.g. BP82
and FP95 for the exact self-similar expressions of these
velocities), the expressions of these velocities are modified by
self-similarity. The slow point is mostly related to the vertical velocity,
the Alfv\'en to the poloidal one, whereas the fast is related to the radial
velocity. 

For given $(\varepsilon, \alpha_m, {\cal P}_m, \chi_m)$, the smooth
crossing of these points is allowed by fine-tuning two parameters, namely
$\mu$ and  $\xi$. For the slow point, the strength of the magnetic field 
$\mu$ must be equal to a critical value $\mu_c$. If $\mu < \mu_c$ the
density profile is flatter, leading to a fall of the vertical speed of
matter, whereas if $\mu > \mu_c$, the magnetic field pinches too strongly the
disc, forbidding matter to leave the disc. The crossing of the Alfv\'en
point is controled by the efficency of ejection $\xi$. If $\xi > \xi_c$ the
magnetic field lines are too much opened because of an overwhelming
centrifugal effect. This leads to $\Omega r^2 > \Omega_* r_A^2$, that is
$B_{\phi} > 0$ and the structure is non-steady. On the contrary, $\xi <
\xi_c$ leads to an unphysical closing of the magnetic surfaces due to the
magnetic tension. For each attempt to cross the Alfv\'en point by
fine-tuning $\xi$, we must find again the critical $\mu$.
\begin{figure}[t]
   \includegraphics[angle=-90,width=\columnwidth]{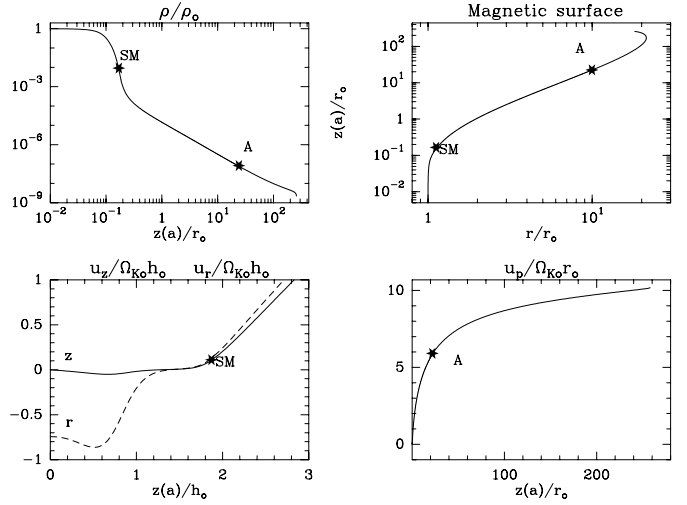}
     \caption{Typical solution from a dissipative disc ($\Lambda$ = 2.43)
       with $\varepsilon = 0.1$, $\alpha_m = 1$, ${\cal P}_m = 3.2$ and
       $\chi_m = 0.259$. All quantities are shown along a magnetic poloidal
       line, anchored on a radius $r_o$. The density (upper left pannel) is
       normalized to the density $\rho_o$ at the disc midplane and jet
       poloidal velocity (lower right pannel) to the Keplerian
       velocity. Stars show the location of the slow-magnetosonic (SM) and
       Alfv\'en (A) critical points. The magnetic surface (upper right
       pannel) presents the usual recollimating behaviour once all MHD
       power has been converted into kinetic power (plateau in the poloidal
       velocity). The lower left pannel shows both radial and vertical
       components of the plasma velocity inside the disc, normalized to the
       sound speed. The transition between the resistive disc and ideal MHD
       jet occurs above the disc surface ($z/h \sim 1.6$), where both
       components become comparable.}
     \label{jetcar}
\end{figure}

As in F97, all trans-Alfv\'enic solutions presented here do not
cross the last critical point, the fast magnetosonic one. The numerical
integration stops at exactly the location of this point (see
Fig.~\ref{jetvel}). The meeting of this last point is unavoidable provided
the numerical integration is done sufficiently far. Actually, it seems
inherent to self-similar solutions that the crossing of the three critical
points is impossible (see discussion in F97).  

Only high-$\omega_A$ jets propagate much farther away than the Alfv\'en
point, widening a lot ($r_{\infty} \gg r_A$) and reaching 
large asymptotic Alfv\'enic Mach numbers ($m_{\infty}^2 \gg 1$). We label
these jets as being powerful. Such jets are produced whenever the initial
MHD Poynting flux is high, namely when the radial current density is large
on the disc midplane. Thus, a high value of $q$ (close to unity) is
required for cold jets launched from a large radial extension in the
disc. Then, such jets exert a large torque on the disc midplane, thereby
producing a large accretion velocity.

\begin{figure*}
  \resizebox{\hsize}{!}{\includegraphics[angle=-90]{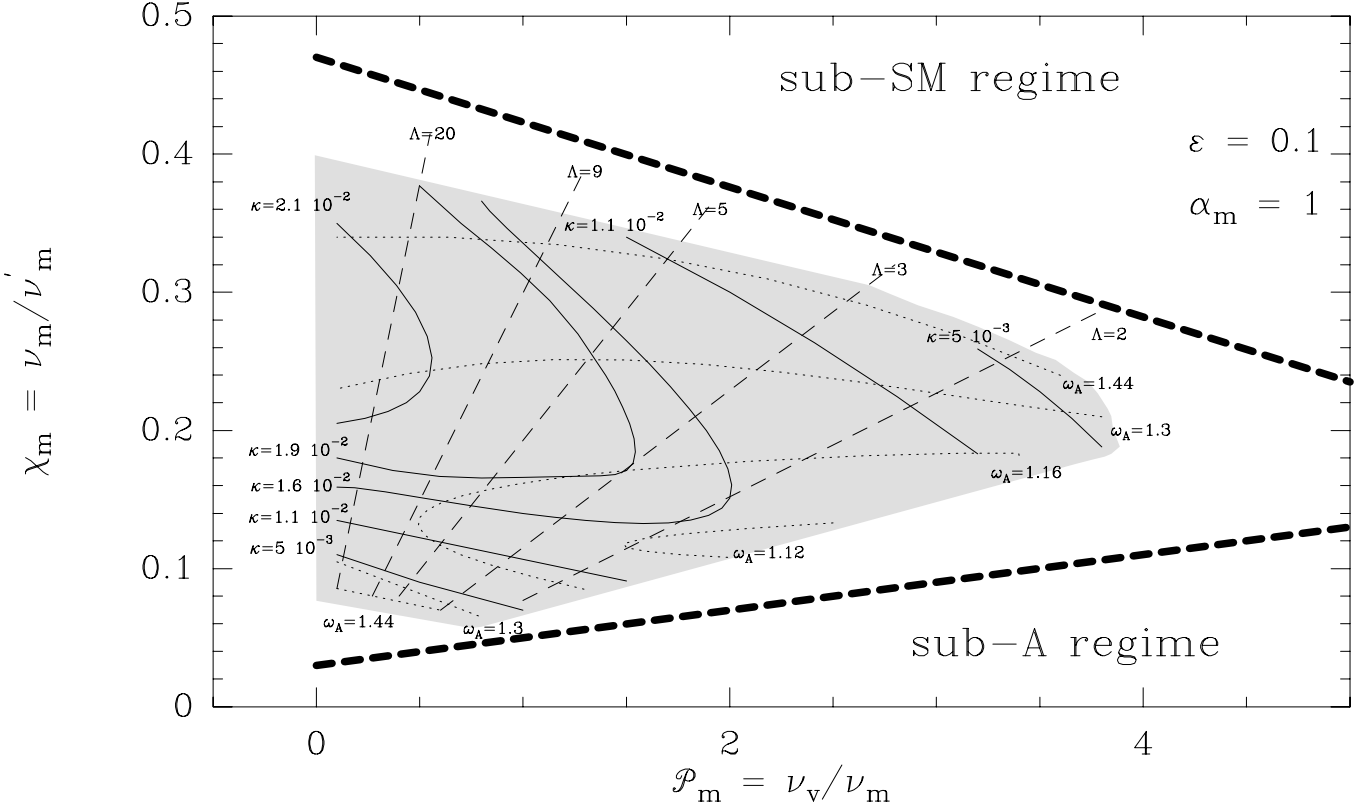}}
  \caption{Parameter space of cold, non-relativistic magnetized
    accretion-ejection structures for $\alpha_m=1$ and
    $\varepsilon=0.1$. The shaded area corresponds to the location where
    numerical solutions could be found. Thick dashed lines show theoretical
    limits: super-slow magnetosonic (upper limit) and super-Alfv\'enic
    (lower limit) flow. While the range in $\chi_m$ is quite narrow, the
    one covered by the magnetic Prandtl number is quite large (from
    $10^{-2}$ to almost 4). Levels for the corresponding jet parameters
    $\kappa$ and $\omega_A$ are also displayed.}
  \label{sp1}
\end{figure*}
For a small poloidal magnetic diffusivity, the poloidal field will tend
to be strongly bent by the accretion flow, eventually providing a large
magnetic Reynolds number. In this case, the turbulent torque becomes
negligible and all the available power goes into the jet. Not much power is
finally radiated at the disc surfaces. Requiring both high-$\omega_A$, cold
jets and a dissipative disc is quite a difficult task, that can however be
achieved in two different (though costly) ways. 

Equation~(\ref{eq:lien}) shows that for a magnetic Prandtl number 
${\cal P}_m$ of order unity there is a strong link between the magnetic
configuration (${\cal R}_m$) and the magnetic torque ($\Lambda$).
Increasing the magnetic diffusivity only ($\alpha_m$), so that
the field lines remain almost vertical inside the disc (${\cal R}_m
\sim 1$), leads to comparable torques. Therefore, a comparable amount of
energy is released as radiation and MHD Poynting flux. The other
possibility consists on increasing the magnetic Prandtl number 
only (${\cal P}_m >1$). In this way, the effects of the turbulent
viscosity are enhanced so that comparable torques can still be achieved, even
with a strongly bent poloidal magnetic field (${\cal R}_m \sim
\varepsilon^{-1}$). 

Actually, it is noteworthy that using the Shakura-Sunyaev prescription for
the viscosity ($\nu_v= \alpha_v \Omega_K h^2$) gives 
\be
\alpha_v = \alpha_m {\cal P}_m \mu^{1/2}\ . 
\ee
\noindent Thus, both ways of obtaining dissipative discs producing
high-$\omega_A$ jets imply an increase of the viscosity parameter
$\alpha_v$. 

\begin{figure*}
  \resizebox{\hsize}{!}{\includegraphics[angle=-90]{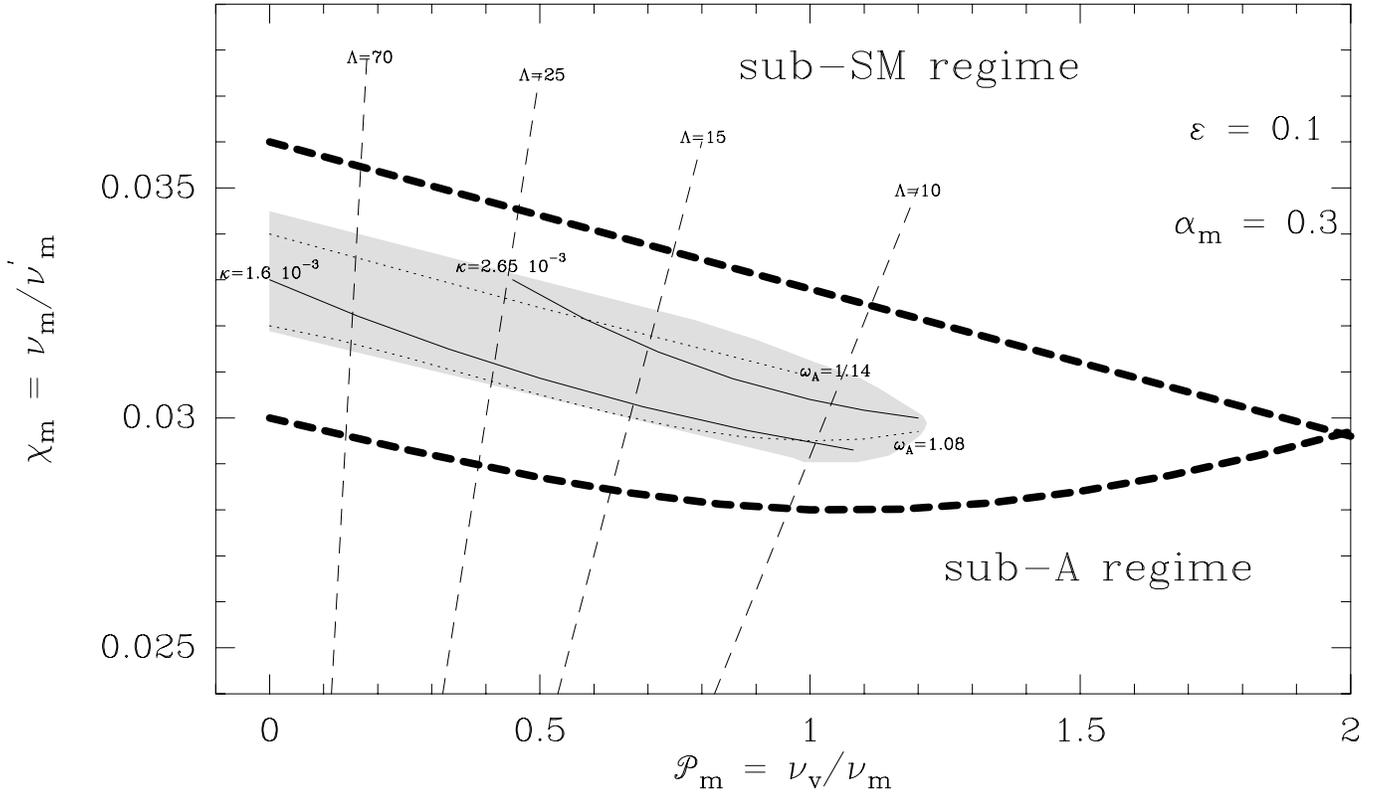}}  
  \caption{Parameter space of cold, non-relativistic magnetized
    accretion-ejection structures for $\alpha_m=0.3$ and
    $\varepsilon=0.1$. For a smaller $\alpha_m$, the region where 
    solutions can be found is much smaller (see Fig.~\ref{sp1}), with
    a magnetic Prandtl number ranging from $10^{-2}$ to 1.2. Here no
    dissipative solution could be found and jets could not reach an
    $\omega_A$ stronger than 1.15.} 
  \label{sp2}
\end{figure*}

Figure~\ref{jetcar} illustrates a typical dissipative solution, described
with the set of parameters $(\varepsilon, \alpha_m,$ $ {\cal P}_m, 
\chi_m$) = (0.1, 1, 3.2, 0.259). This solution is in the upper right side
of the parameter space represented in Fig.~\ref{sp1}. It corresponds to a
dissipative disc ($\Lambda= 2.43$, 29\% of the total mechanical power being
released as radiation) and high-$\omega_A$ jets. The resulting set of jet
parameters is $(\kappa, \lambda, \omega_A) = (5\ 10^{-3}, 78,
1.44)$. 

Inside the disc, both the radial and vertical components of the plasma
velocity are negative. Matter is accreting towards the central object and
slightly converging toward the disc midplane. The transition from accretion
to ejection occurs at the disc surface, where all dynamical terms are
comparable. At the surface, the steep decrease of the density profile 
($\rho^+ = \rho(h) \sim \varepsilon \rho_o$,  where $\rho_o$ is the density
at the disc midplane) goes along with a typical outflow velocity 
\be
u_z^+ \simeq u_o \xi \ .
\ee
\noindent This transition occurs in still resistive and viscous upper
layers, where the magnetic torque azimuthaly accelerates the
plasma. Farther out, both the decrease of the transport coefficients and 
the Lorentz force itself, enforce plasma to flow along magnetic
surfaces. Once in this ideal MHD regime, the flow encounters the first
critical point (SM). 

As previously said, the asymptotic jet behaviour is exactly the same as
in F97. Plasma always achieves its maximum velocity (mostly vertical),
almost all initial MHD Poynting flux being finally converted into kinetic
energy (Fig.~\ref{jetcar}). Jets become super fast-magnetosonic in the
conventionnal sense, namely $u_p > V_{FM} \simeq V_{A\phi}$ ($B_{\phi} \gg
B_p$). From this point on, both the centrifugal force and total pressure
gradient cannot overcome the hoop stress. This leads to a recollimation of
the magnetic surface (negative opening angle), until the last critical
point is finally met (Fig.~\ref{jetvel}). The maximum radius achieved
depends mostly on $\omega_A$ (see Fig.~12 in F97).

High-$\omega_A$ jets from a dissipative disc were obtained here by
increasing the magnetic Prandtl number. In this particular case, we
obtained a phenomenological viscosity parameter $\alpha_v\simeq 2.2$, bigger
than unity. This is problematic since one usually expects a turbulence
which is both subsonic and with a correlation length smaller than the disc
scale height (hence $\alpha_v < 1$). Alternatively, we did find a
dissipative solution with $\alpha_v= 0.27$, but the price was an increase of
the MHD turbulence parameter $\alpha_m = 1.8$. It seems therefore
too costly to get both high-$\omega_A$ jets and dissipative discs. To
firmly settle this issue, we scanned the parameter space of cold MAES.

\begin{figure*}
  \resizebox{\hsize}{!}{\includegraphics[angle=-90]{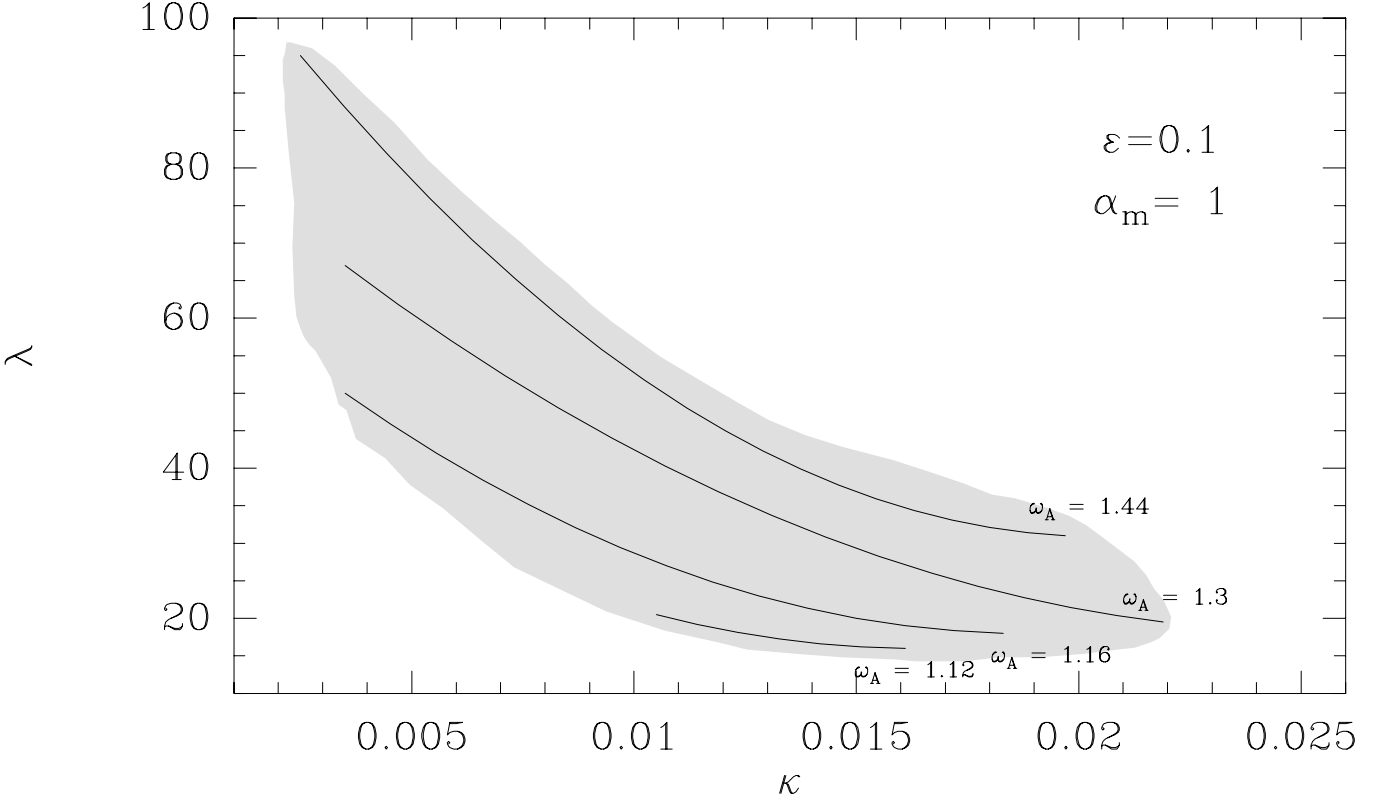}}
  \caption{Parameter space of cold, adiabatic jets from Keplerian discs with
    $\alpha_m=1$ and $\varepsilon=0.1$. It illustrates the disc parameter
    space in Fig.~\ref{sp1}. The lower limit of the shaded area is
    imposed by the Alfv\'enic constraint: no trans-Alfv\'enic jet can be
    found below. The upper limit is imposed by the disc vertical
    equilibrium (trans-SM jet). This parameter space for jets is to be
    compared with those obtained by Blandford \& Payne (\cite{BP82}), Wardle \&
    K\"onigl (\cite{Ward}) and Li (\cite{LiI}). Using this plot along with
    Fig.~\ref{sp1} allows to compute all other MAES parameters ($\xi, \mu,
    q$).} 
  \label{sp3}
\end{figure*}
 
\subsection{Parameter space of adiabatic, cold jets from Keplerian discs}

The parameter space of cold MAES is obtained by varying the set of free
disc parameters ($\varepsilon, \alpha_m, {\cal P}_m, \chi_m$). We choose to
fix the values of both $\varepsilon$ and $\alpha_m$ and represent the
parameter space with the remaining parameters (Fig.~\ref{sp1} and \ref{sp2}).
Numerical solutions are only found inside the shaded areas, where we also
plot levels of two jet parameters $\omega_A$ and $\kappa$. This region is
embedded inside a larger region (thick dashed lines), obtained by two
analytical constraints.  

The first constraint arises from the requirement that jets become
super-slow magnetosonic. It is thus related to the disc vertical
equilibrium: a too strong magnetic pinching of the disc forbids
plasma to escape from it. Since plasma pressure is the only force that
allows matter to be lifted from the disc, it cannot be much smaller than
the vertical component of the magnetic pressure. This magnetic pressure is
due to the growth of both radial and toroidal magnetic fields. Using a
Taylor expansion (on z), this constraint can be written as 
\be 
h\frac{\partial}{\partial z} \left(\frac{B^2_r+B^2_{\phi}}{2
    B^2_o}\right)  \simeq  {\cal R}_m^2\varepsilon^2
\frac{z}{h}\left( 1 +
  \frac{\alpha_m^2\Lambda^2}{4\mu(1+\Lambda)^2}\right) \leq 2 \ ,
\label{compression}
\ee
\noindent Each contribution to the magnetic compression is expected to be
smaller than the one provoked by the vertical magnetic field
(equipartition between plasma pressure and magnetic pressure at the disc
midplane). The upper limit appearing in Fig.~\ref{sp1} and \ref{sp2}  
corresponds to a necessary condition involving only the radial component,
namely   
\be
\chi_m < \frac{\alpha_m^2}{3}(\sqrt{2} - {\cal P}_m \varepsilon) \ .
\ee 
\noindent This Taylor expansion is however not valid in discs where the
poloidal magnetic field is straight (${\cal R}_m \sim 1$). In that case,
there is an extra source of toroidal current at the disc surface, leading
to a large bending of the poloidal field. The corresponding magnetic
pressure increases with the magnetic lever arm. This is what happens in
solutions located at the left lower part of Fig.~\ref{sp1} (high lever arms
$\lambda$ require tiny mass loads $\kappa$, Fig.~\ref{sp3}).

The other constraint emerges from the requirement that jets must become
super-Alfv\'enic. This is expressed by $\omega_A > 1$, which can be written
as 
\be
\chi_m(\chi_m^2-\frac{8\xi\mu}{9\delta}\alpha^2_m) >
\frac{8\alpha_m^4}{27}\frac{\xi\mu\varepsilon}{\delta}{\cal P}_m
\label{transAt}
\ee
\noindent providing the lower limits in the same figures. These two
constraints strongly depend on both $\alpha_m$ and $\varepsilon$.

It turns out that the above two constraints behave differentely with
$\alpha_m$. As a result, the parameter space shrinks considerably (upper
and lower limits merging together) for decreasing $\alpha_m$ and widens for
increasing $\alpha_m$. We show in Fig.~\ref{sp2}, the parameter space for
$\alpha_m$ = 0.3; the space is considerably reduced and produces no
high-$\omega_A$ jet. Physically, it means that if turbulence is not
strong enough, the magnetic torque will not be able to store enough angular
momentum in the magnetic field (parameter $q$) to accelerate matter to the
Alfv\'en surface. A much smaller mass load would avoid this problem, but
this would require an overwhelming magnetic pinching and no trans-SM
solution can be found for cold jets. 
On the other hand, $\alpha_m$ cannot be much bigger than unity otherwise
a too strong magnetic compression due to the toroidal magnetic field will
occur (Eq.~(\ref{compression})). Numerically, we do not found any 
solutions for $\alpha_m > 3$.

No solution with a dominant viscous torque ($\Lambda \ll 1$) has been
found. As showed in Sect.~2.4, super-A jets would require magnetic lever
arms much bigger than $\varepsilon^{-2}$. This is impossible here because
of the tremendous pinching due to the radial component of the magnetic
field. Hence, both super-SM and super-A constraints merge together for
decreasing torque ratios $\Lambda$. Thus, cold disc-driven jets launched
from a large radial extension always carry a large fraction of the disc
angular momentum ($\Lambda >1$). Furthermore, our solutions always produced
a large magnetic lever arm ($\lambda > 3$). This allows us to generalize
the constraint on the maximum ejection index $\xi$ (see Sect.~3.3.2 in
F97), namely
\begin{equation}
\xi < \frac{\sqrt{13}-3}{4}\frac{\Lambda}{\Lambda+1}\ ,
\end{equation}
\noindent which is compatible with the numerical values found.

Only a rather small range of disc aspect ratios allows cold MAES to
exist. This can be understood quite easily. While a maximum value is
imposed by the disc radial equilibrium, cold jet production implies a
minimum value. Indeed, the disc radial equilibrium writes at the midplane
\be
\frac{\Omega_o}{\Omega_K} = \delta = \left( 1 - \varepsilon^2 \left [
    \frac{m_s^2}{2} + 2(2- \beta) + \mu {\cal R}_m \right ] \right)^{1/2}
\label{delta}
\ee
\noindent where 
\begin{eqnarray}
  m_s &\equiv& \frac{u_o}{\Omega_K h} = 2 q \mu \frac{1 + \Lambda}{\Lambda} =
  \alpha_v \varepsilon (1 + \Lambda) \nonumber \\
  &= & \alpha_m ({\cal P}_m\varepsilon + 3\chi_m/\alpha_m^2 )\mu^{1/2}
\end{eqnarray}
\noindent is the sonic Mach number. Deviations from Keplerian law are
mainly due to the radial gradient of plasma pressure (of order
$\varepsilon^2$) and the magnetic tension (of order ${\cal R}_m
\varepsilon^2$). Plasma pressure alone makes it obvious that requiring the
disc to rotate implies that it cannot be too thick. However, the magnetic
tension, which is a slowly decreasing function of height, provides an even
stronger constraint. Indeed, as the density falls down, the magnetic
support becomes more and more important and leads to no rotation (or even a
negative one) for too thick discs. Numerically, we did not found solutions
for $\varepsilon > 0.3$. 

It is noteworthy that there is no dissipative cold MAES for discs 
too thin ($\varepsilon < 0.05$). This is because of the condition for
launching a cold jet. Indeed, Eq.~(\ref{Lambda}) implies that $\chi_m$ must
be of order $\varepsilon$ to have $\Lambda\sim 1$. But this cannot be verified
for very small $\varepsilon$ because of the trans-Alfv\'enic condition
(\ref{transAt}). One can easily see that the thicker disc and the more
dissipative MAES.
  
\begin{figure}[t]
  \resizebox{\hsize}{!}{\includegraphics[angle=-90]{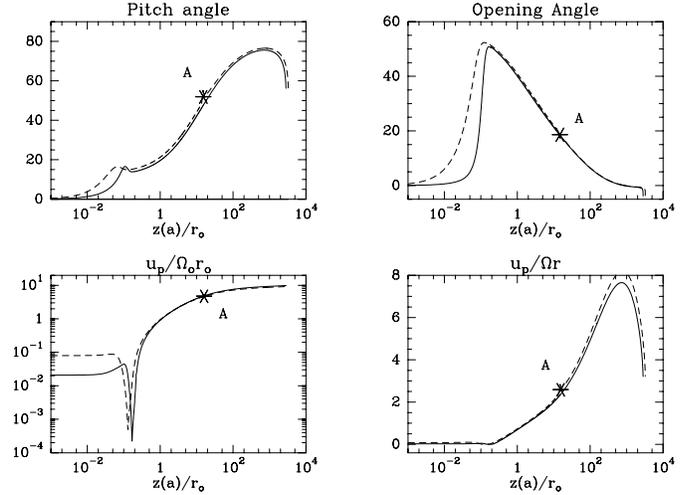}}
  \caption{Plots along a magnetic surface anchored at $r_o$ of the
    principal characteristics of two MAES, with the same set of jet
    parameters but different disc parameters (see Sect. 3.4). One is a
    dissipative disc (solid line) where both torques are comparable
    ($\Lambda= 1.95$), whereas the second is a weakly dissipative disc
    (dashed line) with a much higher magnetic torque ($\Lambda= 10.75$),
    hence a larger accretion velocity. The pitch angle is defined as
    $\arcsin(-B_{\phi}/B_p)$ and the opening angle as $\theta =
    \arccos(B_z/B_p)$. In the ideal MHD region ($z(a)\geq 0.2 r_o$), the
    magnetic field configuration and the velocity field are the same.}
   \label{2jets}
\end{figure}

\subsection{Can identical jets be produced from different discs?}

The complexity of Figures \ref{sp1} and \ref{sp2} raises the question whether
identical jets could be produced from different discs. Once in ideal MHD
regime, equations (\ref{MHDI1}) to (\ref{Grad}) and Eq.~(\ref{polyt})
completely determine the jet structure. The leading equation is the
Grad-Shafranov equation (\ref{Grad}), which provides the jet transverse
equilibrium. In the most general case, the solution is elliptic below the
Alfv\'en surface. This means that the sub-Alfv\'en\-ic solution is completely
determined once we choose the values of all relevant quantities at the
Alfv\'en surface. In the self-similar case, the transfield equation reduces
to a second order differential equation of the form
\be
\frac{d^2r}{dz^2} = \frac{{\cal G}}{{\cal F}}\left(r, \frac{dr}{dz}, z\right)
\ee
\noindent where $r(z)$ describes the shape of magnetic surfaces
(Contopoulos \& Lovelace \cite{Conto}). Therefore, identical jet structures
are produced whenever both the leading dimensionless parameters ($\kappa,
\lambda, \omega_A$) and conditions at the Alfv\'en point are the same. It
can be shown (see Appendix B) that the locus of the Alfv\'en surface is
given by  
\begin{eqnarray} 
  \frac{r_A}{r_o} & = & \lambda^{1/2}  \nonumber \\
   \frac{z_A}{r_A}&= & \cot \phi_A  = \cot\theta_A \left(1 -
    \frac{\omega_A}{\kappa\lambda^{3/2}\cos \theta_A}\right) \ , 
\end{eqnarray}
\noindent where $z_A$ is always found to be of order $r_A$. The 
magnetic configuration (opening and pitch angles) at the Alfv\'en point
satisfies 
\begin{eqnarray} 
  \frac{B_{p,A}}{B_o} &= & \kappa \frac{\lambda^{1/2}}{\omega_A} \nonumber \\ 
  \left . \frac{B_r}{B_z} \right |_A &\equiv & \left .\frac{dr(z)}{dz}
  \right |_A =    \tan \theta_A (\kappa, \lambda, \omega_A) \\ 
  \left . \frac{B_{\phi}}{B_p} \right |_A &= & -g_A \omega_A \nonumber \\
\end{eqnarray}
\noindent whereas dynamical quantities verify
\begin{eqnarray}
& & \rho_A  \frac{\mu_o\Omega_o^2 r_o^2}{B^2_o}=  \kappa^2 \nonumber\\ 
& &  \frac{u_{p,A}}{\Omega_o r_o} =  \frac{\lambda^{1/2}}{\omega_A} \\
& &   \left . \frac{u_p}{\Omega r} \right |_A = \frac{1}{\omega_A (1-g_A)}
  \nonumber 
\end{eqnarray} 
\noindent where both the opening angle $\theta_A$ and the amount $g_A$
of current available at the Alfv\'en surface ($I_A/I_{SM} = g_A(1+ 2
\sigma_{SM}^{-1}) \simeq g_A$) are only functions of $\kappa$, $\lambda$ and
$\omega_A$ (Appendix A). Thus, the three jet parameters impose both the
location of the Alfv\'en surface and the whole jet behaviour.  

As a consequence, if different sets of disc parameters ($\varepsilon$,
$\alpha_m$, ${\cal P}_m$, $\chi_m$) provide the same set of jet parameters
($\kappa, \lambda, \omega_A$), then one gets identical jet configurations from
different discs. The reason why this occurs lies in the resistive MHD
conditions prevailing inside the disc. Thus, as long as there is no theory
fixing the parameters for MHD turbulence, this situation remains. 

\begin{figure*}
   \resizebox{\hsize}{!}{\includegraphics[angle=-90]{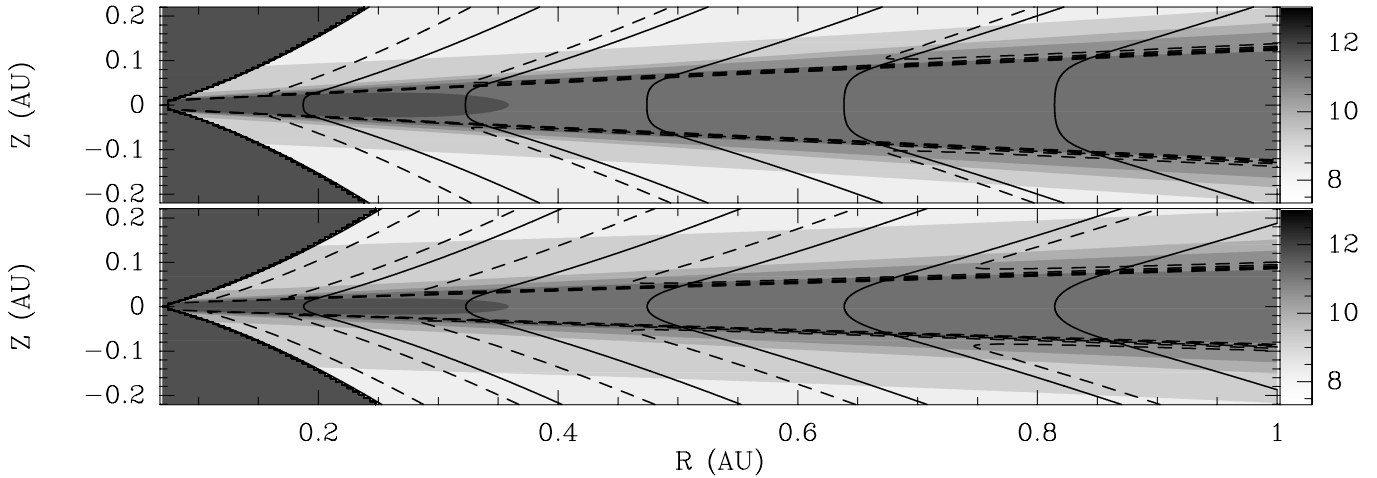}}
     \caption{Poloidal cross-section of two MAES driving jets with the same
       normalized jet parameters as in Fig.~\ref{2jets}, around a one solar
       mass protostar. Poloidal magnetic field lines are displayed in solid
       lines, streamlines in dashed lines and number density ($\log_{10}
       (n_H/\mbox{cm}^{-3}$)) is shown in greyscale, from 12 to 7 by 1.
       In the upper model, the accretion rate, chosen to be $\dot{M}_{ae}=
       10^{-7} M_{\odot}/yr$, is provided by both the turbulent viscous
       torque and the magnetic torque due to the jet. As a result, 34\% of
       the accretion power is released as radiation. In the lower model,
       the magnetic torque due to the jet is dominant (only 8.5\% of the
       power is radiated away), hence providing a higher accretion
       velocity. For the same plasma density at the disc midplane as in the
       upper model, one gets an accretion rate of $3.83\ 10^{-7}
       M_{\odot}/yr$.}
     \label{1jet}
\end{figure*}

We illustrate this property by showing in Fig.~\ref{2jets} the main
characteristics of two jets with roughly (to whithin 10\%) the same
jet parameters. The first one (solid line) is launched from a dissipative
disc with straight magnetic field lines, described by ($\varepsilon,
\alpha_m, {\cal P}_m, \chi_m$) = ($0.1$,$ 1.8$,$ 1$, $0.139$) and releasing 34 \%
($\Lambda = 1.95$) of the total mechanical power. The resulting jets have 
($\kappa$, $\lambda$,$ \omega_A$) = ($7\ 10^{-3}$,$ 65$,$ 1.56$). The
second one (dashed line) is launched from a weakly dissipative disc with
bent magnetic field lines, $(\varepsilon, \alpha_m, {\cal P}_m, \chi_m ) =
(0.1, 0.945, 1, 0.366)$ and releasing only 8.5\% ($\Lambda = 10.75$) of the
mechanical power as radiation. In this case, the resulting set of jet
parameters are almost the same, namely $(\kappa, \lambda, \omega_A) =
(7.8\ 10^{-3}, 58, 1.63)$. However, a computation of a ``real'' MAES
requires more parameters. Indeed, one must also specify the mass $M_*$ of
the central object, both inner ($r_i$) and external ($r_e$) radii where the
MAES is established and the accretion rate $\dot M_{ae}$ at the external
radius. While for most astrophysical objects, the central mass and
accretion rate can be quite severely constrained, the radial extension of a
MAES remains unknown. This is related to the amount of available magnetic
flux, which must be assumed a priori. Thus, if one specifies the same
accretion rate, a larger value of $\Lambda$ (larger magnetic torque)
gives rise to a larger accretion velocity, hence a smaller density at the
disc midplane. This has several consequences that could be used to
discriminate between these two solutions: (1) less dense discs can be
potentially optically thin for the same accretion rate (FP95); (2) the
ratio of the jet kinetic power to the disc luminosity is different; (3)
different absolute values of the disc magnetic field ($\mu$ or $B_o$)
therefore of the jet plasma density ($\rho_A$). In Fig.~\ref{1jet}, we show
the cross sections of two MAES producing the same jets, namely with same
velocities, density stratification and magnetic field configuration. This
was accomplished by imposing the same density at the disc midplane, which
implies a different accretion rate.

To summarize, quasi-identical jet configurations can indeed be obtained from
different discs. However, physical values such as accretion rates (or jet
density) and ratio of disc luminosity to jet kinetic power, would be
different. Thus, there is an one to one correspondance between a disc and
its (cold) jets. 
  
\subsection{Comparison with other steady-state models}

The main assumptions in the work presented here are: (1) discs are
Keplerian and the launching zone is extended, (2) jets are cold. Thus, all
models of jets from accretion discs obtained with these two basic
assumptions should verify our analytical constraints.

Self-similar jets were already obtained by BP82, Wardle \& K\"onigl
(\cite{Ward}) and Li (\cite{LiI}). Basically, our solutions are compatible
with theirs, but we obtained a much smaller parameter space. In particular,
they were able to obtain jets with huge magnetic lever arms $\lambda$ and
correspondingly small mass fluxes $\kappa$. While BP82 did not treat the
disc, Wardle \& K\"onigl (\cite{Ward}) replaced mass conservation with the
prescription $\rho u_z= Constant$ and Li (\cite{LiI}) imposed a static
vertical disc structure. A correct treatment of the disc vertical
equilibrium shows that large magnetic lever arms produce an overwhelming
magnetic pressure squeezing the disc and forbidding any steady-state
solution (see F97 for more details).   

As in BP82, the jets found in this paper always undergo a recollimation,
until the FM critical point is met. This systematic asymptotic behaviour is
strongly influenced by the chosen self-similar geometry
(F97). Nevertheless, using exactly the same self-similar ansatz,
Contopoulos \& Lovelace (\cite{Conto}) and more recently Ostriker
(\cite{Eve}) did found jets with different behaviours. They were able to
find jets with either an oscillating pattern or reaching a cylindrical
collimation. This is possible only by varying the magnetic flux
distribution in the disc, namely for $\beta > 1$. However, disc-driven jets
must follow the scaling $\beta= 3/4 + \xi/2$ (FP95) where the
trans-Alfv\'enic constraint imposes $\xi < 1/2$. Thus cold jets with $\beta
> 1$ do not match the boundary conditions imposed by Keplerian accretion discs.

To our knowledge, only Li (\cite{LiII}) produced magnetically-driven jets
from a Keplerian disc with straight magnetic field lines inside. In his model,
ambipolar diffusion is the source of diffusivity in a quasi-neutral disc,
whereas accretion is solely produced by the magnetic torque due to the
jet (no turbulent ``viscous'' torque). Thus, almost no power is radiated by
the disc and ${\cal P}_m << 1$. However, his solution displayed $\mu <<1$,
thereby allowing a high mass loss rate from the disc (weak magnetic
compression). This is outside our parameter range where we found $\mu$ to
be always of order unity. This discrepancy might arise from the different
set of equations used. In particular, Li (\cite{LiII}) did not use the
induction equation for the magnetic field, but assumed instead frozen-in
ions and got the magnetic field behaviour through the prescribed vertical
dependency of the coupling coefficient.

\section{Summary and concluding remarks}

We investigated the full parameter space of cold, non-relativis\-tic,
magnetized accretion ejection structures (MAES). A generalization of
previous works (FP95, F97) has been made by taking into
account a radial transport of angular momentum in the disc. This turbulent
transport is assumed to arise from MHD instabilities triggered inside the
disc. As a consequence, a cold MAES is characterized by four local parameters:
the disc aspect ratio $\varepsilon$, the strength of the MHD turbulence
$\alpha_m$, the magnetic Prandtl number ${\cal P}_m$ and the turbulent
dissipation aniso\-tropy $\chi_m$. 

Using self-similar solutions, we were able to obtain continuous solutions,
from the disc midplane to super-Alfv\'enic jets. The parameter range where
cold solutions are possible is quite narrow: $10^{-3} \leq \varepsilon
<0.3$, $0.3 \leq \alpha_m < 3$, $ 0 \leq {\cal P}_m \la 4$ and $10^{-3}<
\chi_m < 1$. Inside this parameter space, both dissipative and
weakly-dissipative discs are allowed. A dissipative disc releases a
significant fraction of the accretion (mechanical) power as radiation at
its surfaces. The magnetic torque is comparable to the turbulent
``viscous'' torque ($\Lambda \sim 1$). A weakly dissipative disc transfers
most of the available power to an MHD Poynting flux, which is finally
converted into jet kinetic power. Only a small fraction of the power is
radiated away, and the magnetic torque is dominant ($\Lambda \gg 1$). 

The importance of the dissipation is controlled by the properties of the
MHD turbulence. But since a MAES must also magnetically drive cold winds,
there is a link between these properties and jet launching conditions. This
crucial relation is expressed in Eq.~(\ref{Lambda}), namely
\be
\Lambda \sim \frac{\chi_m/ \varepsilon}{\alpha_m \alpha_v} \ .
\ee
\noindent Dissipative discs require $\chi_m \sim \varepsilon$,
i.e. anisotropic magnetic diffusivities for thin discs. In other words,
turbulence must lead to a dissipation of the toroidal field stronger than
for the poloidal field. Dissipative MAES require either $\alpha_m$ or
$\alpha_v$ bigger than unity. Thus, if one demands those parameters to be
always smal\-ler than unity, only weakly dissipative MAES are possible. We
believe this is a general result for steady-state structures. It 
arose from analytical considerations and is thus independent of our
self-similar ansatz. However, it does depend on two assumptions, namely a
Keplerian disc and cold ejection. 

Applying these assumptions to an extended launching zone, we derived
general analytical links between relevant jet parameters ($\kappa, \lambda$
and $\omega_A$) and disc parameters. The most powerful relation, linking
accretion to ejection, is Eq.~(\ref{qlamb})
\be
\omega_A \sim m_s
\frac{\Lambda}{1+\Lambda}\frac{\lambda^{3/2}}{\lambda-1}\frac{\sin(
  \phi_A-\theta_A)}{\sin \phi_A}
\ee
\noindent which must be bigger than (but of the order of) unity. The larger
$\omega_A$, the larger the maximum jet radius. The above equation, coming
from magnetic flux conservation, shows how this jet parameter is intimately
related to the accretion process ($m_s= u_o/\Omega_K h$). This is quite
obvious for a dominant magnetic torque, where the toroidal field at the
disc surface is large. Would the ``viscous'' torque be dominant ($\Lambda
\ll 1$), the sonic Mach number would be very small, $m_s \sim \alpha_v
\varepsilon$. The constraint $\omega_A > 1$ would then require a huge
magnetic lever arm from a thin disc. It is doubtful such solutions could
exist, the disc being probably too much squeezed by the magnetic pinching
force. Another alternative would be a tremendously high viscosity parameter
$\alpha_v \sim \varepsilon^{-1}$, which is unphysical. 

Two biases were nevertheless introduced in our numerical solutions. The
first one arises from the geometry itself. Indeed, self-similar solutions
are only valid for MAES settled on a large radial extension. As a 
consequence, a cylindrical rather than a spherical geometry is already
imposed. This fixes the Alfv\'en surface, so that $z_A \sim r_A$ is always
verified ($\phi_A$ between $20^{\circ}$ and $50^{\circ}$), hence making gravity
negligible there. As a result, it can be shown that $\Lambda$ must be
larger than (or of the order of) unity. Therefore, no cold solution with
$\Lambda \ll 1$ on a large radial extension can be found. However, note
that such a small value of $\Lambda$ would have tremendous implications on
the MHD turbulence parameters $\chi_m$, $\alpha_m$ and $\alpha_v$.

The second bias was already found by PP92. A class of jets described with
the same parameters in all magnetic surfaces become super-fast magnetosonic
(in the conventional sense, $u_p > V_{FM}$) and then recollimate toward the
axis. Self-similar jets from Keplerian discs fall into this category
(F97). A realistic modelling of self-collimated jets must therefore be done
outside a self-similar framework. However, the interrelations between the
underlying disc and its jets (links between invariants) should be
used. Ouyed et al. (\cite{ouyed}) performed numerical simulations of MHD
jets from Keplerian accretion discs, using the disc as a boundary
condition. They obtained solutions that recollimate also toward the axis
and produce fast MHD shocks. Knots are episodically formed at a fixed
position and, once formed, propagate further down the jet. They proposed
that this recollimation and unsteady knot production is realized whenever
the toroidal magnetic field is large enough, namely when $N=
B_{\phi}^2/2\mu_o \rho u_p^2 > 1$ at the disc surface. It is remarkable
that our calculations of cold MAES satisfy this criterion.

We showed that there is an univocal link between an accretion disc and its
jets. But interpreting observations is quite tricky, since it depends on
assumptions about jet emission properties (Cabrit et
al. \cite{Cabrit}). However, global considerations can already provide some
constraints on theoretical models, like the ratios $2\dot M_j/\dot M_a$ ($=
\xi \ln r_e/r_i$) and $P_{jet}/P_{rad}$ ($= \Lambda$). If $\Lambda$ is
found to be larger than unity, then the portion of the disc responsible for
mass ejection (from $r_i$ to $r_e$) should consistently produce a hole in
the spectral energy distribution. Moreover, since we found an ejection
efficiency $\xi$ which lies typically around $0.01$, mass fluxes could
provide another measure of the MAES's radial extension. Young stellar
objects produce jets with typical velocities of 300 to 500 km/s. If we
assume that these jets are produced by a MAES located after the corotation
radius, too high velocities are obtained. This is because the magnetic
lever arm is quite large. Thus, models with smaller lever arms $\lambda$
should be sought. However, this is impossible for cold jets within a
self-similar framework. 

Another alternative would be that the emission is dominated by the external
radii of the MAES. In that case, only the outer velocities would be
inferred  from the observational data, thus underestimating asymptotic
velocities. To settle this question, one needs to solve the energy equation
along the jet and produce synthetic emission maps (Garcia et al., in
preparation).  
 
As said previously, we trust our results are quite general for Keplerian
accretion discs launching cold jets from a large radial
extension. The next generalization is to relax the assumption of cold
ejection. Indeed, even if plasma pressure may have no dynamical role in jet
equilibrium, it might produce non-negligible effects on the ejection
mechanism itself. In particular, since the vertical equilibrium at the disc
surface may be deeply modified, different mass loads and/or initial jet
velocities are expected. The presence of hot coronae is quite commonly
argued for accretion discs, from both theoretical (Galeev et
al. \cite{Galeev}) and observational (e.g. Kwan \cite{Kwan})
grounds. Introducing a hot corona is consistent with the presence of a
turbulent ``viscous'' torque. Indeed, such a torque probably arises from 3D
MHD turbulence in the disc. Numerical simulations tend to show the
formation of a magnetized, hot medium above the disc (Stone et
al. \cite{stone}). In our model, a small fraction of the mechanical power
liberated by this torque could be converted into coronal heating. This work
is currently under progress. 

Finally, it has been argued that advection dominated accretion flows
could also produce self-collimated jets ( see Narayan \& Yi \cite{narayan},
Soria et al. \cite{soria}). These are quite promising configurations,
especially in the context of galactic and extragalactic compact objects. In
this paper, we mainly focused our attention on thin (even slim) accretion
discs. But our framework, with the further development of a hot corona,
will allow us to investigate the behaviour of flows of that kind.   

\begin{acknowledgements}
We would like to thank Guy Pelletier for stimulating discussions and 
helpful comments about the manuscript. 
\end{acknowledgements}

\appendix
\section{Physical constraints at the Alfv\'en point}

In Grad-Shafranov equation (\ref{Grad}) evaluated at the Alfv\'en point
($m=1$), the most difficult term to figure out is the density gradient. To
solve this problem, we consider the function $g=1 - \Omega/\Omega_*$, which
can be written   
\be
 g  = \frac{m^2}{m^2 - 1 }\left( 1 - \frac{r^2_A}{r^2}\right) \ .
\ee
\noindent Taking the gradient of this equation and noting that $m^2 =
\rho_A/\rho$ provides
\begin{eqnarray}
\nabla\ln \left(\frac{g}{\rho_A}\right) & = &\frac{1}{\rho_A-\rho}
\left(\frac{2r_A}{r^2}(\nabla r - \nabla r_A)\right.\nonumber\\ &+&
\left.\frac{}{}\nabla\rho - \nabla\rho_A\right)
\end{eqnarray}
\noindent Thus, the regularity condition at the Alfv\'en point is 
\be 
\left.\nabla\ln\rho\right|_A = \nabla\ln\rho_A + \frac{2}{g_Ar_A}(\nabla
r_A - {\bf e}_r)
\label{rhoa}
\ee 
\noindent where we made the assumption that jet invariants are the same
in every magnetic surface. Inserting the regularity condition in
the Grad-Shafranov equation (\ref{Grad}) provides
\begin{eqnarray}
  \cos \theta_A & = &\frac{g_Ar^2_A\eta^2}{2B_{p,A}} \left[ 
    \frac{d{\cal E}}{da} + (1-g_A)\Omega_*r^2_A\frac{d\Omega_*}{da} + 
    g_A\Omega_*\frac{d\Omega_* r^2_A}{da} \right. \nonumber\\ 
& + &  \left.  \frac{2V^2_{Ap,A}}{g_Ar_A} \frac{dr_A}{da} +
  \frac{V^2_{Ap,A}}{2}(1 + \omega^2_Ag^2_A ) \frac{d\ln\rho_A}{da} \right]
\end{eqnarray}
This is a general result as long as the jet parameters are the same for
every magnetic surface. In the case of radial self-similarity, it takes the
much simpler form
\begin{eqnarray}
  \cos \theta_A & = &  \frac{g_A}{2} \frac{\kappa\lambda^{3/2}}{\omega_A}
  \frac{\Omega_o}{\Omega_*} \left[\omega^2_A\left(\frac{\alpha_4}{2}g^2_A +
    2g_A+\frac{3}{2} \left \{\frac{1}{\lambda}-1 \right\} \right)
  \right. \nonumber\\   
  & + & \left.\frac{\alpha_4}{2}+ \frac{2}{g_A}\right]
\label{eq:GS}
\end{eqnarray}
\noindent where $\alpha_4= 2\beta - 3$. Thus, the opening angle at the
Alfv\'en point depends only on $g_A$, $\kappa$, $\lambda$ and $\omega_A$.
Now, Bernoulli equation evaluated at the Alfv\'en surface provides
\be 
g^2_A = 1 -\frac{3}{\lambda} -
\frac{1}{\omega^2_A}+ \frac{2}{\lambda^{3/2}(1+z^2_A/r^2_A)^{1/2}} \ .  
\label{eq:BB}
\ee
\noindent Note that the last expression is valid for every model of
disc-driven, cold jets. If the magnetic lever arm is very small ($\lambda <
3$), then the fastness parameter must be large, namely $\omega_A > 2$, even
taking into account the gravity term ($g^2_A >0$).

\section{Shape of the Alfv\'{e}n surface and $\omega_A$}

The definition (\ref{defome}) of $\omega_A$ can be written, using the
definition of the mass load $\kappa$, 
\begin{equation}
  \label{defome2}
  \omega_A =
  \kappa\lambda^{1/2}\sqrt{\frac{\Omega_o}{\Omega_*}}\frac{B_o}{B_{p,A}} \
  . 
\end{equation}
\noindent If we assume the Alfv\'{e}n surface to be conical, as in radial
self-similar works or in some numerical simulations of disc-driven jets
(Sakurai \cite{sak}, Krasnopolsky et al. \cite{Kras}), then the magnetic flux conservation provides
\begin{equation}
  \label{magnecons}
  \frac{B_{o}}{B_{p,A}} = \frac{\Omega_o}{\Omega_*}\lambda\frac{\sin
  (\phi_A - \theta_A)}{\sin \phi_A}  
\end{equation}
\noindent where $\phi_A$ is the angle between the Alfv\'{e}n surface and
the vertical axis ($z_A/r_A = \cot \phi_A$ in a self-similar
solution). Injecting this result in Eq.~(\ref{defome2}) gives 
\begin{equation}
  \label{defome3}
  \omega_A \simeq \kappa\lambda^{3/2}\frac{\sin (\phi_A - \theta_A)}{\sin
  \phi_A}  \ ,
\end{equation}
\noindent since $\Omega_o \simeq \Omega_*$ in thin discs. Our solutions always
displayed $ 1 < \omega_A < 2$ and never higher values. The reason for that
behaviour is hidden in Grad-Shafranov equation (\ref{eq:GS}), which imposes
a highly non-linear link between $\omega_A$ and the opening angle
$\theta_A$. Nevertheless, we observe that $\theta_A$ increases with
$\omega_A$ (see e.g. Fig.~6 in F97), while the above expression of
$\omega_A$ shows that it should decrease for increasing $\theta_A$. Thus,
the jet transverse equilibrium seems to impose here some feedback,
forbidding high values for $\omega_A$. As a consequence, Bernoulli 
equation prohibits small values of the magnetic lever arm $\lambda$.

Smaller magnetic lever arms could however be obtained for different
geometries. To illustrate this, we examine a spherical Alfv\'{e}n surface,
in the particular case where magnetic field lines are almost straight
(Najita \& Shu \cite{Naj}). In this case, one has  
\begin{eqnarray}
  \label{spher}
  \frac{B^+}{B_{p,A}} & \simeq & \frac{\lambda}{\varepsilon^2 \sin^2
 \theta_A} \nonumber \\
  \omega_A &\simeq&
  \kappa\frac{\lambda^{3/2}}{\varepsilon^2\sin^2\theta_A}\frac{B_o}{B^+}
\end{eqnarray}
\noindent where $B^+$ is the poloidal magnetic field at the disc
surface. Large values of $\omega_A$ with small magnetic lever arms are
allowed here because of the large dilution of the magnetic flux.


\begin{thebibliography}{}
\bibitem[1991]{Balb} Balbus, A.S., Hawley, J.F., 1991, ApJ, 376, 214 
\bibitem[1982]{BP82} Blandford, R.D., Payne, D.G., 1982, MNRAS, 199, 883 (BP82)
\bibitem[1999]{Cabrit} Cabrit, S., Ferreira, J., Raga, A.C., 1999, A\&A,
  343, L61  
\bibitem[1980]{Chan} Chan, K.L., Henriksen, R.N., 1980, ApJ, 241, 534
\bibitem[1994]{Conto} Contopoulos, J., Lovelace, R.V.E., 1994, ApJ, 429, 139 
\bibitem[1993]{FP93} Ferreira, J., Pelletier, G., 1993, A\&A, 276, 625 (FP93)
\bibitem[1995]{FP95} Ferreira, J., Pelletier, G., 1995, A\&A, 295, 807 (FP95)
\bibitem[1997]{F97} Ferreira, J., 1997, A\&A, 319, 340 (F97)
\bibitem[1979]{Galeev} Galeev, A.A., Rosner, R. , Vaiana, G.S., 1979, ApJ,
  229, 318 
\bibitem[1995]{Har}  Hartigan, P., Edwards, S., Ghandour, L., 1995, ApJ,
  452, 736 
\bibitem[1989]{Hey} Heyvaerts, J.,
  Norman, C., 1989, ApJ, 347, 1055 
\bibitem[1994]{Khan} Khanna, R., Camenzind, M., 1994, ApJ, 435, L129 
\bibitem[1989]{Ko} K\"onigl, A., 1989, ApJ, 342, 208 
\bibitem[1997]{Kwan} Kwan, J., 1997, ApJ, 489, 284
\bibitem[1998]{lery} 
Lery, T., Heyvaerts, J., Appl, S., Norman, C.A., 1998, A\&A, 337, 603      
\bibitem[1999]{LeryII} Lery, T., Henricksen, R.N., Fiege, J.D., 1999, A\&A,
  350, 254 
\bibitem[1995]{LiI} Li, Z.-Y., 1995, ApJ, 444, 848
\bibitem[1996]{LiII} Li, Z.-Y., 1996, ApJ, 465, 855
\bibitem[1997]{Livio} Livio, M., 1997 ASP conference Series, Vol 121,
  Wickramasinghe, T., Ferrario, L., Bicknel, G.V. (eds.), p 845
\bibitem[1999]{Kras} Krasnopolsky, R., Li, Z.Y., Blandford, R., to appear
  in ApJ (Dec 1999). Draft version at astro-ph/9902200
\bibitem[1969]{Michel} Michel, F.C., 1969, ApJ, 158, 727 
\bibitem[1994]{Naj} Najita, J., Shu, F.H. 1994, ApJ, 429, 808 
\bibitem[1995]{narayan} Narayan, R., Yi, I., 1995, ApJ, 444, 231
\bibitem[1997]{Eve} Ostriker, E.C., 1997, ApJ, 486, 291
\bibitem[1997]{ouyed} Ouyed, R., Pudritz, R.E., Stone, J.M., 1997, Nature,
  385, 409 
\bibitem[1992]{PP92} Pelletier, G., Pudritz, R.E., 1992, ApJ, 394, 117 (PP92)
\bibitem[1976]{pouq76} 
Pouquet, A., Frisch, U., Leorat, J., 1976, J. Fluid Mech., 77, 321 
\bibitem[1981]{Pudr} Pudritz, R.E., 1981, MNRAS, 195, 897 
\bibitem[1995]{Rud} R\"{u}diger, G., Elstner, D., Stepinski, T.F., 1995,
  A\&A, 298, 934  
\bibitem[1992]{rutten} Rutten, R.G.M., van Paradijs, J., Tinbergen, J.,
  1992, A\&A, 260, 213 
\bibitem[1987]{sak} Sakurai, T., 1987, PASJ, 39, 821
\bibitem[1998]{Sarj} Serjeant S., Rawlings, S., Lacy, M., Maddox, S.J.,
  Baker, J.C., Clements, D., Lilje, P.B., 1998, MNRAS, 294, 494 
\bibitem[1973]{Shak} Shakura, N.I., Sunyaev, R.A., 1973, A\&A, 24, 337 
\bibitem[1997]{soria} Soria, R., Li, J., Wickramasinghe, D. T., 1997, ApJ,
  487, 769 
\bibitem[1996]{stone} Stone, J.M., Hawley, J.F., Gammie, C.F., Balbus,
  A.S., 1996, ApJ, 463, 656 
\bibitem[1994]{Brand} Torkelsson, U., Brandenburg, A., 1994, A\&A, 283, 677
\bibitem[1993]{Ward} Wardle, M., K\"onigl, A., 1993, ApJ, 410, 218
\bibitem[1993]{Yosh} Yoshizawa, A., Yokoi, N., 1993, ApJ, 407, 540
\end{thebibliography}
\end{document}